\documentclass[12pt]{article}
\begin{document}
\title{EFFECT OF  MAGNETIC FIELD ON THE PHASE TRANSITION FROM NUCLEAR MATTER TO QUARK MATTER DURING PROTO-NEUTRON STAR EVOLUTION}
\author{V.K.Gupta \footnote{E--mail : vkg@ducos.ernet.in}, Asha Gupta, S.Singh\footnote{E--mail : santokh@ducos.ernet.in}, J.D.Anand\footnote{E--mail : jda@ducos.ernet.in} \\
 	{\em Department of Physics and Astrophysics,} \\
	{\em University of Delhi, Delhi-110 007, India.} \\
        {\em InterUniversity Centre for Astronomy and Astrophysics,} \\
        {\em Ganeshkhind, Pune 411007 , India.} \\
        }
\renewcommand{\today}{}
\setlength\textwidth{5.75 in}
\setlength\topmargin{-1.cm}
\setlength\textheight{8 in}
\addtolength\evensidemargin{-1.cm}
\addtolength\oddsidemargin{-1.cm}
\font\tenrm=cmr10
\def\baselinestretch{1.4}

\maketitle
\large
\begin{abstract}
We have studied phase transition from hadron matter to quark matter in the presence of high magnetic fields incorporating the trapped electron neutrinos at finite temperatures. We have used the density dependent quark mass (DDQM) model for the quark phase while the hadron phase is treated in the frame-work of relativistic mean field theory. It is seen that the nuclear energy at phase transition decreases with both magnetic field and temperature. A brief discussion of the effect of magnetic field in supernova explosions and proto-neutron star evolution is given.
\end{abstract}
\pagebreak
\begin{section} {Introduction}

The phase transition from hadron matter to quark matter (QM) is expected to occur in a variety of physically different situations. In the astrophysical context this transition can take place in the interior of a cold, extremely dense neutron star. The other and the more likely situation in which such a transition can take place is the proto-neutron star (Prakash, Cooke, and Lattimer 1995; Prakash et al. 1997). These proto-neutron stars (PNS) have very high temperatures in their interiors ($\sim$50MeV) and more importantly, have a large amount of trapped neutrinos (Burrows, and Lattimer 1986; Bethe 1990; Keil and Janka 1995). It is well known that the structure of a PNS changes significantly in a few seconds after its formation as a result of the release of most of its energy and lepton excess by means of neutrino emission. After this PNS stage, the evolution of the neutron star (NS) may be described basically as a cooling process in which the thermal and structural properties are almost decoupled.
It is worthwhile to look for signals of quark deconfinement during the short PNS evolution time scale, which is intimately related to the supernova event itself. Recently a lot of studies of the dynamical properties of core collapse on the effect of the transition to QM have been undertaken (Mezzacappa et al. 1998; Janka 1996). Such studies were motivated by the difficulties of the standard theoritical models in accounting for the observed explosions. In one of the recent studies Lugones and Benvenuto (1998) have considered the transition from nuclear matter to QM under conditions prevailing in a PNS during the first tens of seconds since its birth. They considered a hyperonic equation of state (EOS) for the nuclear phase and the `bag model' EOS for the QM phase. They studied the effect of the mass of the strange quark, $m_s$, and the bag constant, B, on the phase transition.\\
\\
With the discovery of intense outbursts of low energy $\gamma$-rays from neutron stars there has been a lot of interest in the study of NS and their intense magnetic fields. Kuoveliotou $\b{et}$ $\b{al}$ (1998) have recently found a soft $\gamma$-ray repeater SGR 1806-20 with a period of 7.47s and a spin down rate of 2.6$\times10^{-3} syr^{-1}$ from which they estimate the pulsar's age to be about 1500 years and field strength of $\sim$ 8$\times10^{14}$G. Since most models of magnetic field generation give core fields to be $10^{3}$- $10^{5}$ times higher than the surface fields, the magnetic field in the core of such neutron stars, called magnetars, could be as high as $10^{18}$- $10^{19}$ G or even $10^{20}$G. Such intense magnetic fields may not be ruled out by the virial theorem if one takes significant general relativistic corrections into account (Hong 1998). Further Kuoveliotou $\b{et}$ $\b{al}$ (1998) have estimated that upto 10$\%$ of all known NS could be these magnetars. Since it is well known that observationally it is very difficult to distinguish normal neutron stars from strange stars (SS) or hybrids (neutron stars with strange matter (SM) cores), some, if not all, of the observed intense magnetic field pulsars could be these exotic objects, formed by the deconfinement transition from nuclear matter to quark matter during the PNS stage itself under an intense magnetic field.\\
\\
Since such super strong magnetic fields affect the EOS (Anand and Singh 1999) and almost all known properties of NS or SS quite significantly, it will be interesting to study their effect on the deconfinement transition itself.
Motivated by this, we in this paper study the viability of the transition from nuclear matter to quark matter in the conditions prevailing in such a magnetized PNS during the first few seconds of its birth. We follow closely the approach of Lugones and Benvenuto (1998, 1999). For the description of nuclear matter phase we have employed the well known mean field theory developed by Walecka, Serot, Glendenning and others (See Glendenning 1997 and references therein). We have included contribution of $\sigma$, $\omega$, $\rho$ interactions as well as scalar self interaction terms; with these one obtains the correct experimental values of the various physical constants relevant in nuclear matter, viz, binding energy/nucleon, the saturation nuclear density, the asymmetry energy, the bulk modulus and the effective mass. Matter in this hadronic phase consists of neutrons, protons, electrons $\&$ electron neutrinos and their antiparticles; however we have not included the muon neutrino or the full baryon octet in this treatment.\\
\\
For the description of the QM phase, consisting of ud quarks, we have used the alternative description of QM in which the confinement is treated by assuming a baryon density dependence of quark mass introduced by Fowler, Raha and Weiner(1981) and Plumber, Raha and Weiner(1984). Later on it was reformulated to show that the properties of SM in this density dependent quark mass model are quite similar to those predicted by the bag model and yet it is much easier to handle at finite temperatures. Recently we have used this DDQM model, in conjuction with strong magnetic fields, to study the radial pulsations of strange stars ( Anand et al. 2000).\\
\\ 
In section 2 a brief discussion of the EOS for the nuclear phase and the EOS for the QM at finite temperatures and in the presence of a magnetic field is given. Also included in section 2 is the criterion for finding  phase transition from nuclear phase to QM phase described by Lugones and Benvenuto (1998). Section 3 deals with results and discussion.
\end{section}

\vspace {0.5cm}
\begin{section} {The transition from nuclear matter to quark matter} 

2.1 The equation of state for nuclear matter\\
\\
Following Lugonese and Benvenuto (1998), we apply the nuclear mean field model for nuclear matter EOS. We incorporate n, p, e and $\bar{\nu_e}$ particles and their antiparticles (since the antiparticles are also present at a finite temperature).
In the absense of magnetic field, expressions for the number, pressure and energies densities are 
\begin{equation}
n_i = \frac{g_i}{(2\pi)^3} \int d^3p[f_{i}(T)-\bar{f_i}(T)]
\end{equation}
\begin{equation}
P_i =\frac{1}{3} \frac{g_i}{(2\pi)^3} \int \frac{p^2d^3p[f_{i}(T)+\bar{f_i}(T)]}{(p^2 +m_i^{2})^{1/2}}
\end{equation}
\begin{equation}
\rho_i = \frac{g_i}{(2\pi)^3} \int  d^3p(p^2 +m_i^{2})^{1/2} [f_{i}(T)+\bar{f_i}(T)]
\end{equation}
where i=n, p, e, $\nu_e$. The spin degenracy factor, $g_i$, is unity for $\nu_e$ and two for all others. Here $f_i(T)$ and $\bar{f_i}(T)$ are the fermi-dirac distribution functions for particles and antiparticles respectively:
\begin{equation}
f_i(T) = (\exp([\sqrt{p^2+m_i^{2}} - \mu_i]/T)+1)^{-1}
\end{equation}
\begin{equation}
\bar{f_i}(T) = (\exp([\sqrt{p^2+m_i^{2}} + \mu_i]/T)+1)^{-1}
\end{equation}
These expressions are valid for all the particles; however for neutron and proton, as a result of the strong interactions as described by the mean field theory, the mass $m_i$ has to be replaced by the effective mass $m_i^*$ and the chemical potential $\mu_i$ by the effective chemical potential $\mu_i^*$:
\begin{equation}
m_i^* = m_i -  g_\sigma\sigma
\end{equation}
\begin{eqnarray}
\mu_i^* = \mu_i  - \left(\frac{g_\omega}{m_\omega}\right)^2~ \rho_B^{'} - I_{3Z}\left(\frac{g_\rho}{m_\rho}\right)^{2}~\rho_{I_3}^{'} 
\end{eqnarray}
The iso-spin density $\rho_{I_3}^{'}$ and the baryon densities $\rho_B^{'}$ are given by
\begin{equation}
\rho_{I_3}^{'} =\frac{1}{2} ( \rho_p -\rho_{n} )
\end{equation}
\begin{equation}
\rho_{B}^{'}  = \rho_p +\rho_n 
\end{equation}
The total pressure P and the mass energy density $\rho$ for the system are given by 
\begin{eqnarray}
P &=& \Sigma P_i + \frac{1}{2}\left(\frac{g_\omega}{m_\omega}\right)^2~ \rho_{B}^{'2} - \frac{1}{2}\left(\frac{g_\sigma}{m_\sigma}\right)^{-2} (g_\sigma~\sigma)^2 - \frac{1}{3}~b~m_N~(g_\sigma~\sigma)^3 \nonumber \\ 
&-& \frac{1}{4}~c~(g_\sigma~\sigma)^4 + \frac{1}{2}\left(\frac{g_\rho}{m_\rho}\right)^2~\rho_{I_3}^{'2} 
\end{eqnarray}
\begin{eqnarray}
\rho &=& \Sigma \rho_i + \frac{1}{2}\left(\frac{g_\omega}{m_\omega}\right)^2~ \rho_{B}^{'2} + \frac{1}{2}\left(\frac{g_\sigma}{m_\sigma}\right)^{-2} (g_\sigma~\sigma)^2 + \frac{1}{3}~b~m_N~(g_\sigma~\sigma)^3 \nonumber \\ 
&+& \frac{1}{4}~c~(g_\sigma~\sigma)^4 + \frac{1}{2}\left(\frac{g_\rho}{m_\rho}\right)^2~\rho_{I_3}^{'2} 
\end{eqnarray}
where (Glendenning, 1997) 
\begin{eqnarray}
\left(\frac{g_\sigma}{m_\sigma}\right)^2 = 11.79~ fm^{-2} &,& \left(\frac{g_\omega}{m_\omega}\right)^2 = 7.149 fm^{-2} \nonumber \\
\left(\frac{g_\rho}{m_\rho}\right)^2 = 4.411 fm^{-2} &,& b = 0.002947, ~c = -0.001070
\end{eqnarray}
The mean field $g_\sigma\sigma$ satisfies
\begin{equation}
\left(\frac{g_\sigma}{m_\sigma}\right)^{-2} (g_\sigma~\sigma) + b~m_N~(g_\sigma~\sigma)^2 +c~(g_\sigma~\sigma)^3  = \Sigma n_i^s
\end{equation}
Here $g_\sigma$, $g_\omega$, $g_\rho$, $m_\sigma$, $m_\omega$, $m_\rho$ are the sigma, omega, rho couplings and masses respectively, b and c are the strengths of the two scalar self interaction terms in the Lagrangian:
\begin{eqnarray}
L &=& \sum_{i=n,p}{\bar{\psi_i}} [i\gamma_{\mu}(\partial^{\mu} + ig_{\omega}\omega^{\mu} + \frac{i}{2}g_{\rho}\vec{\tau}~.\vec{\rho}~^{\mu}) - (m -g_\sigma~\sigma)]\psi_i \nonumber \\
& & + \frac{1}{2} ( \partial_{\mu} \sigma \partial^{\mu} \sigma - m_{\sigma}^{2}~\sigma^{2}~)-\frac{1}{4} \omega_{\mu\nu}\omega^{\mu\nu} +\frac{1}{2}m_{\omega}^{2}\omega_{\mu}\omega^{\mu} \nonumber \\
& & -\frac{1}{4}\vec{\rho}_{\mu\nu}~\vec{\rho}~^{\mu\nu} +\frac{1}{2}m_{\rho}^{2}\vec{\rho}_{\mu}.\vec{\rho}^{\mu} -\frac{1}{3} bm(g_\sigma~\sigma)^{3} \nonumber \\
& & -\frac{1}{4}c (g_\sigma~\sigma)^{4}   
\end{eqnarray}
In addition to the number density $n_i$, for neutron and proton we also require the scalar density $n_i^s$:
\begin{equation}
n_i^s = \frac{g_i}{(2\pi)^3} \int \frac{m_i^*d^3p[f_{i}(T)+\bar{f_i}(T)]}{(p^2 +m_i^{*2})^{1/2}}
\end{equation}
We now introduce a constant and uniform magnetic field of magnitude B in the positive Z direction. The motion of a charged particle of charge e and mass m in such a magnetic field gets quantized and the energy eigenvalues are given by \\
$~~~~~~~~~~~~~~~E^2 = p_z^2 +m^2 +2(n +1/2 + s)eB$\\ 
where n=0, 1, 2, 3,..... defines the various Landau levels and s=$\pm1/2$ refers to the spin up (down) states of the particle. The number of states per unit volume in an interval of longitudinal momentum $\triangle p_z$ for a given level n is $(2-\delta_{no})\frac{eB}{(2\pi)^2}\triangle p_z$. [we are using $\hbar$ = c = 1]. Hence the usual sum over particle states (per unit volume) in zero field :\\
$~~~~~~~~~~~~~~~~~~~~~~~~~\frac{2}{(2\pi)^3} \int d^3p$ \\  
(2 being the spin degenracy factor) must be replaced by \\
$~~~~~~~~~~~~~~~~~~~~~~~~~\frac{eB}{2\pi^2}\sum_{n}(2-\delta_{no}) \int dp_z$\\
With this ansatz it is straightforward to obtain expressions for the energy density $\rho$, pressure density P, baryon density n and the scalar density $n^s$ of a charged particle in magnetic field. These expressions are:
\begin{equation}
n_i = \frac{g_i eB}{2\pi^2}\sum_{n}(2-\delta_{no}) \int dp_{z}[f_{i}(T)-\bar{f_i}(T)]
\end{equation}
\begin{equation}
n_i^s = \frac{g_i eBm_i^*}{2\pi^2}\sum_{n}(2-\delta_{no}) \int \frac{dp_{z}[f_{i}(T)+\bar{f_i}(T)]}{E_i^*}
\end{equation}
\begin{equation}
P_i = \frac{g_i eB}{2\pi^2}\sum_{n}(2-\delta_{no}) \int \frac{p_z^2dp_{z}[f_{i}(T)+\bar{f_i}(T)]}{E_{i}^{*}}
\end{equation}
\begin{equation}
\rho_i = \frac{g_i eB}{2\pi^2}\sum_{n}(2-\delta_{no}) \int E_i dp_{z}(f_{i}(T)+\bar{f_i}(T))
\end{equation}
\\
The hadron phase is assumed to be charge neutral and in $\beta$-equilibrium. Electric charge neutrality requires
\begin{equation}
n_p = n_e 
\end{equation}
Chemical $\beta$-equilibrium in the presence of trapped electron neutrinos implies the following relation between the chemical potentials of various particles involved in the hadron phase: 
\begin{equation}
\mu_p = \mu_n -( \mu_e - \mu_{\nu_e})
\end{equation}
All the above equations can be solved numerically by giving four quantities, the temperature T, the electron chemical potential $\mu_{e}$, the chemical potentials of the electron neutrinos $\mu_{\nu_e}$ and the magnetic field B.\\
\\
2.2 The equation of state for quark matter \\
\\
The quark phase is composed of u, d quarks, gluons, electrons and electron neutrinos. We shall treat QM as a free fermi gas of u, d quarks where the mass of each quark is parametrized as
\begin{equation}
m_u = m_d =\frac{C}{3n_B},
\end{equation}
C being a  constant. The range of C is to be constrained by stability arguments. The pressure in the quark phase is given by 
\begin{equation}
P = \Sigma P_i
\end{equation}
where the sum goes over i = u, d, e, $\nu_e$ and their antiparticles. For the electron and the electron-nentrino, pressure is given by eq(2) in the absence of magnetic field and by eq(18) in the presence of magnetic field. To obtain expressions for P, $\rho$ and n for u and d quarks, we use the thermodynamic potential which in the presence of magnetic field, B, is given by
\begin{eqnarray}
\Omega_i &=& - \frac{g_ie_{i}BT}{2\pi^2}\sum_{n}(2-\delta_{no}) \int dp_{z}[\log(1+\exp(-\beta[E_i^* -\mu_i])) \nonumber \\
 &+&\log(1+\exp(-\beta[E_i^* +\mu_i]))]
\end{eqnarray}
where 
\begin{equation}
E_i^* = \sqrt{p_z^2 +m_i^2 +2nq_iB}
\end{equation}
$g_i$= 3 (colour) and $q_i$= (2/3,-1/3) for u and d quark respectively.
The pressure $P_i$ is given by (Peng et al. 2000) 
\begin{equation}
P_i = n_B \frac{\partial\Omega_i}{\partial n_B} -\Omega_i
\end{equation}
energy density $\rho_i$ by
\begin{equation}
\rho_i =\Omega_i +\mu_in_i -T \frac{\partial\Omega_i}{\partial T} 
\end{equation}
and number density $n_i$ by
\begin{equation}
n_i = - \frac{\partial\Omega_i}{\partial\mu_i} 
\end{equation}
Using eq(24) for the thermodynamic potential $\Omega_i$, yields:
\begin{eqnarray}
P_i &=& \frac{g_i e_{i}B}{2\pi^2}\sum_{n}(2-\delta_{no}) \int \frac{p_z^2dp_{z}[f_{i}(T)+\bar{f_i}(T)]}{E_{i}^*} \nonumber \\
&-&\frac{C}{3 n_{B}} \frac{g_i e_{i}Bm_{i} }{2\pi^2}\sum_{n}(2-\delta_{no}) \int \frac{dp_{z}[f_{i}(T)+\bar{f_i}(T)]}{E_{i}^*}
\end{eqnarray}
The expressions for number density and energy density remain the same as in hadron phase.\\
\\
2.3 Transition from PNS to quark matter \\
\\
Following Benvenuto $\b{et}$ $\b{al}$ it is assumed that the transition from nuclear matter to quark matter in chemical equilibrium occurs necessarily through an intermediate step of quark deconfinement driven by strong interactions. In view of this, in a deconfinement transition the abundance per baryon of each quark and lepton flavour is the same in both the hadron and quark phases. This is in contrast to the model first  given by Glendenning (1992) who proposed the existence of a mixed phase.\\
\\
In order to compute the conditions for phase transition, we apply the standard Gibb's criteria i.e, equality of pressure P, temperature T and Gibb's energy per baryon g in both phases: 
\begin{equation}
P_q = P_h ~,       ~~~  T_q=T_h ~,  ~~~   g_q=g_{h}
\end{equation}
together with flavour per baryon coservation equations
\begin{equation}
Y_i^q = Y_i^h
\end{equation}
for i=u, d, e, $\nu_e$. Notice that this condition automatically makes the quark phase to be charge neutral. Here
\begin{equation}
Y_i^x = \frac{n_i}{n_B^x}             
\end{equation}
for  x =  h or q. Also
\begin{eqnarray}
Y_u^h & = & 2Y_p +Y_n \nonumber \\
Y_d^h & = & Y_p + 2Y_n
\end{eqnarray}
Starting from $Y_p$ and $Y_n$, obtained from the hadronic phase, one then obtains various number densities and thereby the chemical potentials in the quark phase; energy and pressure are then obtained from Eqs(19) and (29) respectively.
\end{section}

\vspace {0.5cm}
\begin{section} { Results and discussion } 
In this work we are interested in the deconfinement transition from hadron to quark matter in the presence of trapped neutrinos and magnetic field. As discussed in detail by Benvenuto and Lugonese, we see that for a fixed set of values of $\mu_{\nu_e}$, T and eB the nuclear phase depends only on one of the chemical potentials (say $\mu_p$). With $\mu_p$, $\mu_e$, T and eB fixed, we can calculate $n_p$, $n_n$, $n_e$, $n_{\nu_e}$ and so the abundances $Y_p$, $Y_n$, $Y_e$, $Y_{\nu_e}$, $Y_u^h$, $Y_d^h$. Using conditions (30), (31) we obtain the mass energy density U of hadron matter at which it deconfines.
At finite temperature, we have to evaluate numerically Fermi-Dirac integrals appearing in $n_i$, $n_{i}^{s}$, $P_i$ and $E_i$. We were able to evaluate these integrals upto an accuracy of about 0.1$\%$ by using a technique due to Cloutman (1989).
At this stage it is necessary to explain this method by means of a test integral. Consider the following integral occuring in $n_i$:
\begin{eqnarray}
I_i&=& \int_{-\infty}^{+\infty} dp_z f_i(T) \nonumber \\
   &=&2\int_{0}^{\infty}\frac{dp_z}{1+e^{\beta(E_i^*-\mu_i^*)}} 
\end{eqnarray}
Here
\begin{equation}
E_i^*= (p_z^2 +m_i^{*2})^{1/2}
\end{equation}
Putting $\beta p_z$ =x, we have
\begin{eqnarray}
I_i =2T\int_{0}^{\infty}\frac{dx}{1+e^{(x^2+m_i^{'2})^{1/2}-\mu_i^{'}}}
\end{eqnarray}
where
\begin{eqnarray}
m_i^{'}&=& m_i^{*} \beta \nonumber \\
 \mu_i^{'}&=& \mu_i^{*}\beta
\end{eqnarray}
We can write another form for $I_i$ by substituting
\begin{eqnarray}
t=(x^2+m_i^{'2})^{1/2}-m_i^{'}
\end{eqnarray}
viz.,
\begin{eqnarray}
I_i &=&2T\int_{0}^{\infty}\frac{(t+m_{i}^{'})dt}{[\sqrt{t(t+2m_{i}^{'})}][1+e^{t-\eta}]} \nonumber \\
    &=&2T\int_{0}^{\infty}\frac{\phi^{'}(t)}{[1+e^{t-\eta}]}dt 
\end{eqnarray}
where 
\begin{eqnarray}
\eta&=&\mu_i^{'} -m_i^{'} \nonumber \\
  \phi^{'}(t)&=&\frac{t+m_{i}^{'}}{\sqrt{t(t+2m_{i}^{'})}}
\end{eqnarray}
Depending upon the values of $\eta$ we can have various forms of $I_i$: \\
\\
Case 1: $\eta>30$
\begin{equation}
I_i = \phi(\eta) + 2\sum_{j=1}^{5} C_{2j} \phi^{2j}
\end{equation}
where
\begin{eqnarray}
\phi(\eta) &=& \int_{0}^{\eta}\phi(t) dt \nonumber \\
\phi^{2j}&=& \frac{\partial^{2j-1}\phi^{'}}{\partial t^{2j-1}}
\end{eqnarray}
Case 2: -5 $\leq$ $\eta$ $\leq$ 30 \\
The original form (eq 34) is good for evaluating $I_i$ numerically by Simpson's rule. The upper limit and the number of points were adjusted in such a way that the evaluation of integral was accurate to about 0.1$\%$. \\
\\
Case 3: $\eta <-5 $  \\
We expand $I_i$ in terms of modified bessel function $K_1$.
\begin{equation}
I_i \simeq \sum_{\nu=0}^{\infty} (-1)^{\nu}m_i^{'}  e^{\mu_{i}^{'}(\nu +1)} K_{1}(m_{i}^{'}(\nu +1))
\end{equation}
Using the above method of evaluating integrals at finite temperatures, the transition from PNS to QM can be determined at a given temperature for various values of $\mu_{\nu_e}$ provided the parameter C is known.
We choose C in such a way that the ud system is unstable. We have chosen three sets of values of C namely 90, 110, 130 MeV$fm^{-3}$. \\
\\
In fig 1, we have plotted the mass energy density, U, of hadron matter (in terms of the nuclear saturation density, $\rho_o$ = 2.7 $\times 10^{14}$gm$cm^{-3}$) at which phase transition occurs verses the chemical potential of neutrino present in hadron matter for three values of C, viz, C=90, 110, 130 MeV$fm^{-3}$ and at temperatures T=5 and 60 MeV. We find that the phase transition density is an increasing function of the neutrino chemical potential. It is interesting to note that the phase transition density increases with increase in the value of C. On the other hand, the energy density is a decreasing function of temperature T. Our results in this regard are similar to those of Lugones and Benvenuto (1998) who have used the bag model EOS for the quark system. In both cases the energy density required for the transition to take place increases with increasing $\mu_{\nu_e}$, but for given $\mu_{\nu_e}$ decreases with increasing temperature. In quantitative terms the energy density at transition in our model is less by about a factor of two as compared to the Lugones-Benvenuto results.
In fig2 we have studied the phase transition of hadron matter to quark matter in the presence of a strong magnetic field. We have taken magnetic fields of 0, 5$\times$$10^4$ and $10^5$ $MeV^2$ (1$~MeV^2$ $\sim$ 1.6$\times10^{14}$G) at temperatures 5 and 60 MeV and C=90 MeV$fm^{-3}$. The results are shown as a graph of U vs. $\mu_{\nu_e}$. It is interesting to note that the phase transition density decreases with increase of magnetic field at a given temperature. Similar results are obtained for higher C as well at a given temperature.
Figure 3 deals with the behaviour of phase transition density with temperature for different values of $\mu_{\nu_e}$. The  phase transition density decreases from 2.2$\rho_o$ to about 1.3$\rho_o$ for eB=0 when $\mu_{\nu_e}$ varies from 300 MeV to 0 MeV, whereas for eB=$10^5$$MeV^2$ U falls from about 2$\rho_o$ to 0.8$\rho_o$. 
Figure 4 shows the same kind of results for C=110 MeV$fm^{-3}$. \\
\\
To conclude, we find the phase transition density is lowered in the presence of magnetic field. In this regard the effect of magnetic field is similar to that of temperature. The presence of a strong magnetic field is likely to cause a transition to free quark matter in the proto-neutron stars in early stages of its evolution. The question can however be settled by studying the EOS and M-R relation for such a magnetized PNS to see whether it meets the conditions of the energy density required for the transition to take place. This question is under study and will be reported elsewhere.   
\end{section}

\pagebreak

Figure captions
\vskip 0.5 cm
Figure 1. The mass energy density, U, of hadron matter at which phase transition occurs versus the chemical potential of the electron neutrinos present at eB=0. Labels a1, a2 and a3 correspond to T=5 MeV and C=90, 110 and 130 MeV$fm^{-3}$ respectively, whereas b1, b2 and b3 correspond to T=60 MeV and C=90, 110 and 130 MeV$fm^{-3}$ respectively.    
\vskip 0.5 cm
Figure 2. The mass energy density, U, of hadron matter at which phase transition occurs versus the chemical potential of the electron neutrinos present at C=90 MeV$fm^{-3}$. Labels a1, a2 correspond to eB =0 and T=5 and 60 MeV respectively; b1, b2 correspond to eB=5$\times$$10^4 MeV^2$ (1$~MeV^2$ $\sim$ 1.6$\times10^{14}$G), T=5 and 60 MeV respectively and c1, c2 correspond to eB= $10^5 MeV^2$ T=5 and 60 MeV respectively.    
\vskip 0.5 cm
Figure 3. The hadron matter mass energy density of phase transition, U, versus the temperature T at C=90 MeV$fm^{-3}$. Labels a0, a1, a2, a3  correspond to eB=0 and $\mu_{\nu_e}$=0, 100, 200, 300 MeV respectively; b0, b1, b2, b3  correspond to eB=$10^5 MeV^2$ and $\mu_{\nu_e}$=0, 100, 200, 300 MeV respectively.
\vskip 0.5 cm
Figure 4. The same as fig.3 but for C=110 MeV$fm^{-3}$. 
\pagebreak

\begin{figure}[ht]
\vskip 15truecm
\includegraphics{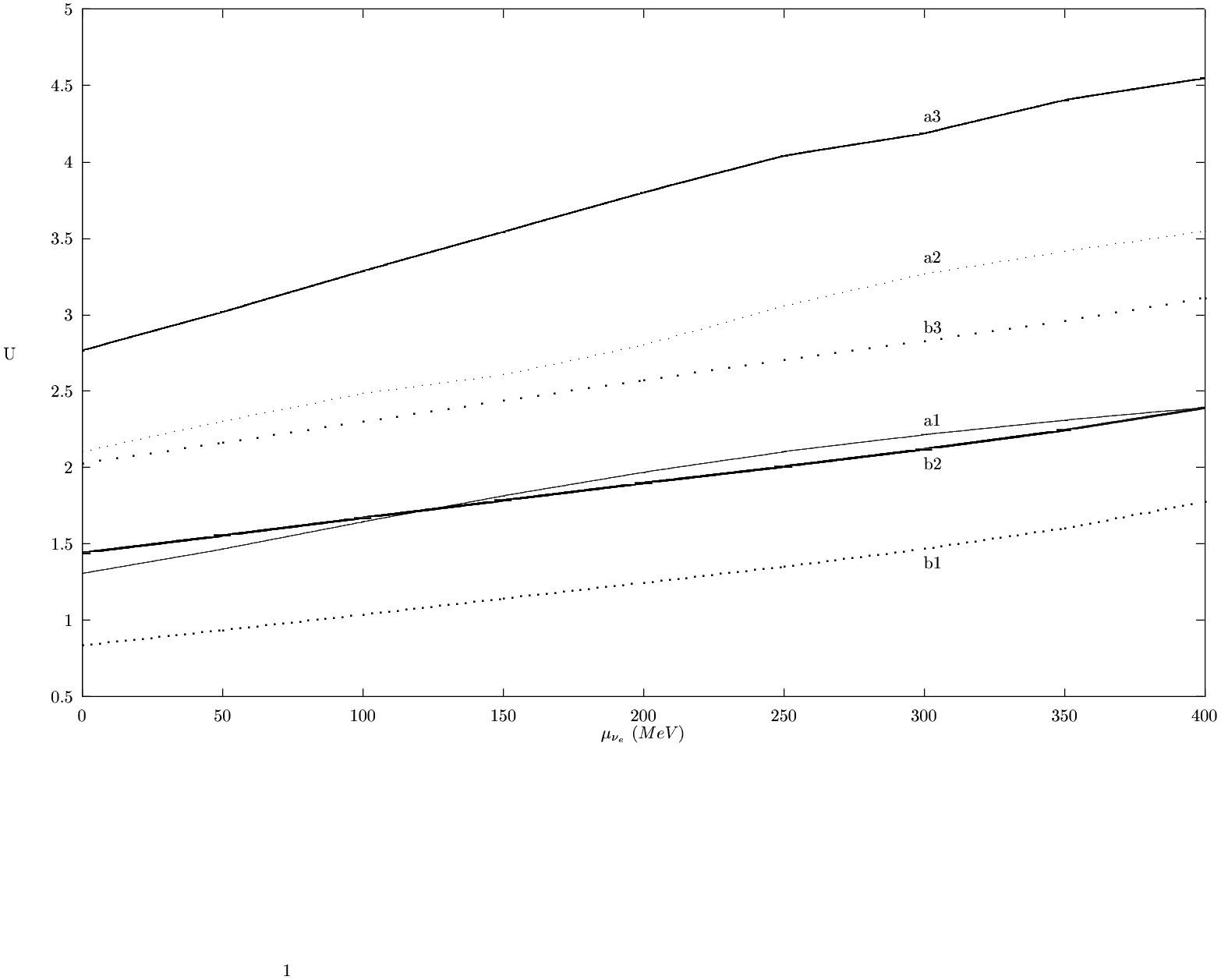}
\caption{The mass energy density, U, of hadron matter at which phase transition occurs versus the chemical potential of the electron neutrinos present at eB=0. Labels a1, a2 and a3 correspond to T=5 MeV and C=90, 110 and 130 MeV$fm^{-3}$ respectively, whereas b1, b2 and b3 correspond to T=60 MeV and C=90, 110 and 130 MeV$fm^{-3}$ respectively.}
\end{figure}

\begin{figure}[ht]
\vskip 15truecm
\includegraphics{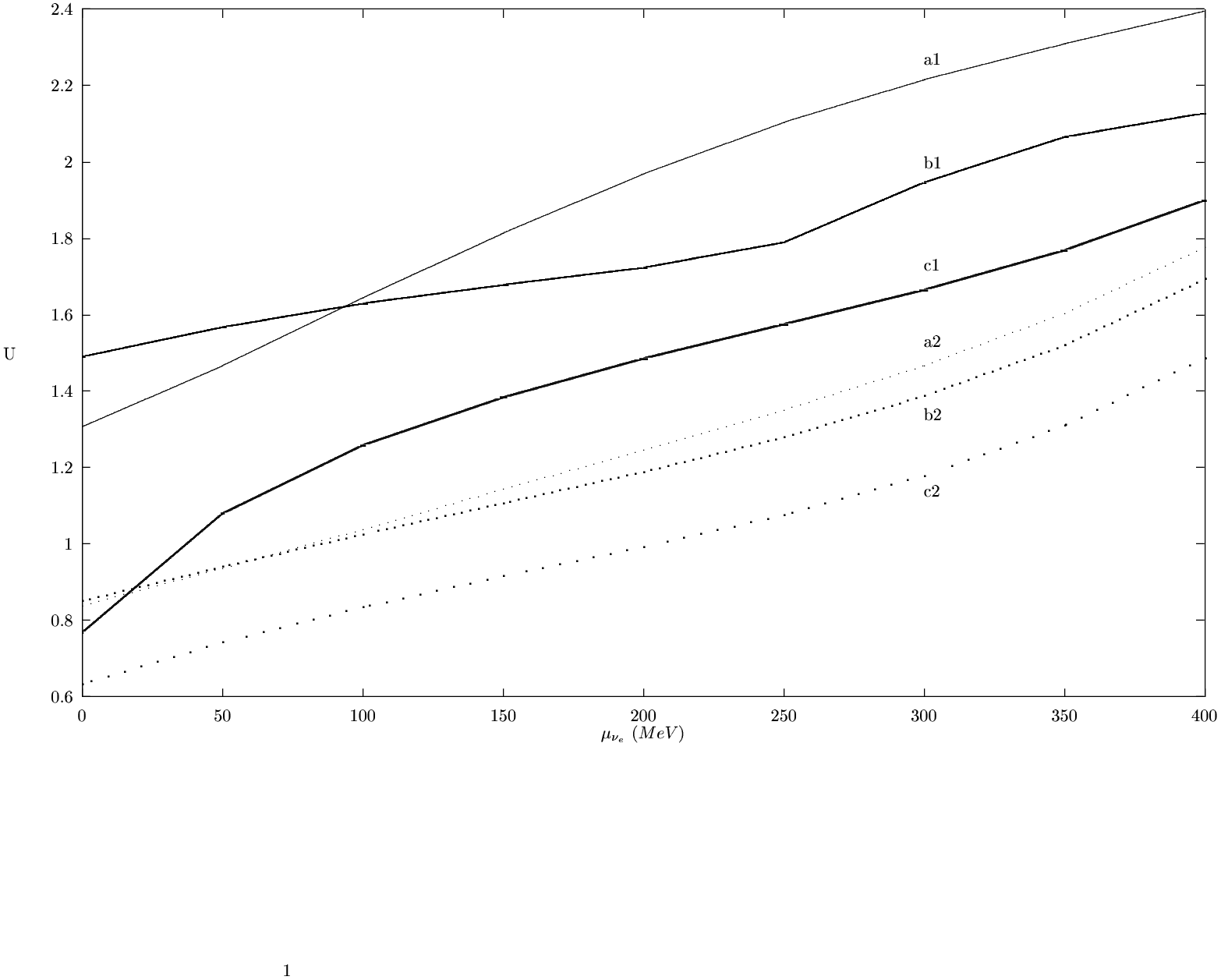}
\caption{The mass energy density, U, of hadron matter at which phase transition occurs versus the chemical potential of the electron neutrinos present at C=90 MeV$fm^{-3}$. Labels a1, a2 correspond to eB =0 and T=5 and 60 MeV respectively; b1, b2 correspond to eB=5$\times$$10^4 MeV^2$ (1$~MeV^2$ $\sim$ 1.6$\times10^{14}$G), T=5 and 60 MeV respectively and c1, c2 correspond to eB= $10^5 MeV^2$ T=5 and 60 MeV respectively.}
\end{figure}

\begin{figure}[ht]
\vskip 15truecm
\includegraphics{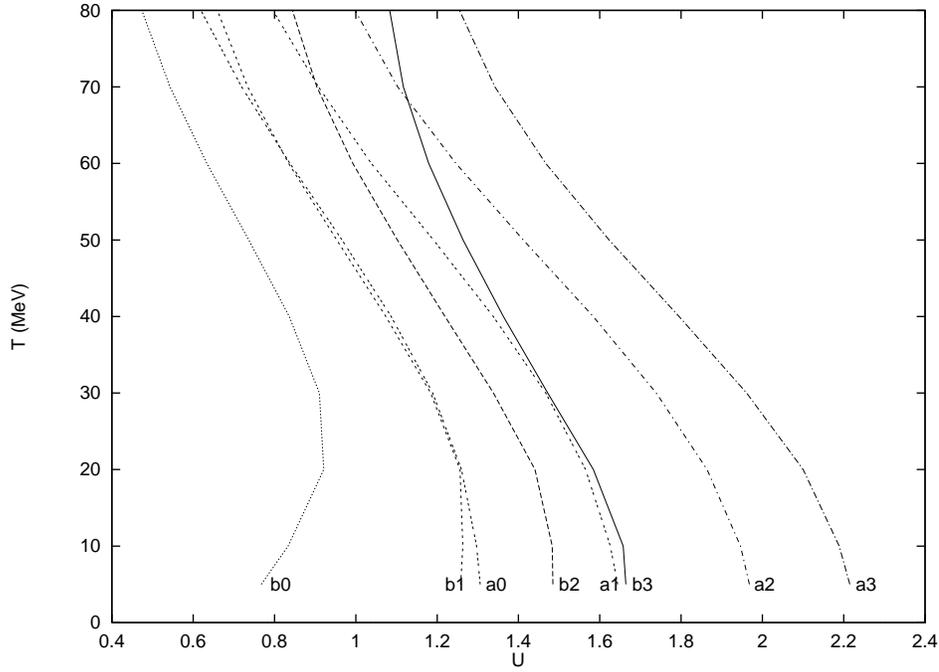}
\caption{The hadron matter mass energy density of phase transition, U, versus the temperature T at C=90 MeV$fm^{-3}$. Labels a0, a1, a2, a3  correspond to eB=0 and $\mu_{\nu_e}$=0, 100, 200, 300 MeV respectively; b0, b1, b2, b3  correspond to eB=$10^5 MeV^2$ and $\mu_{\nu_e}$=0, 100, 200, 300 MeV respectively.}
\end{figure}

\begin{figure}[ht]
\vskip 15truecm
\includegraphics{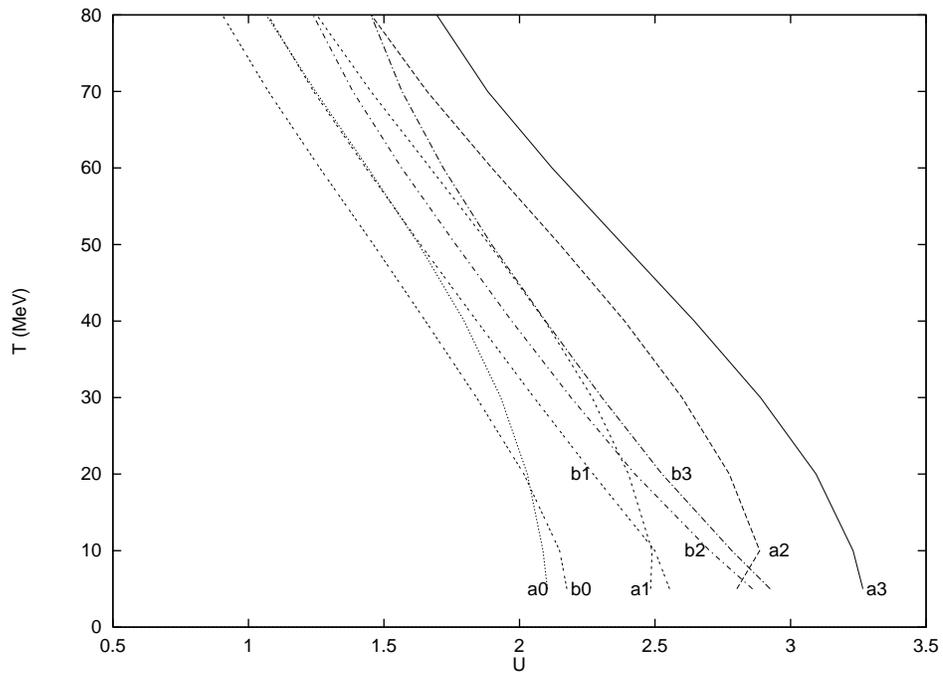}
\caption{The same as fig.3 but for C=110 MeV$fm^{-3}$.}
\end{figure}

\end{document}